\documentclass[journal=accacs,manuscript=article,layout=twocolumn]{achemso}
\usepackage{chemformula} 
\usepackage[T1]{fontenc} 
\usepackage[utf8]{inputenc}
\usepackage{graphicx}
\usepackage{epsfig}
\usepackage{epstopdf}
\usepackage{float}
\usepackage{amsmath}
\usepackage{caption}
\usepackage{subcaption}
\usepackage{natbib}
\usepackage{xcolor}
\usepackage{lineno}
\usepackage{braket}
\usepackage{multirow}
\usepackage{amssymb}
\usepackage[colorlinks=true,linkcolor=blue]{hyperref}

\author{Manoj Dey}
\affiliation{Materials Research Centre, Indian Institute of Science, Bangalore 560012, India}
\author{Ritesh Kumar}
\affiliation{Materials Research Centre, Indian Institute of Science, Bangalore 560012, India}
\author{Abhishek Kumar Singh}
\affiliation{Materials Research Centre, Indian Institute of Science, Bangalore 560012, India}
\email{abhishek@iisc.ac.in}

\title[An \textsf{achemso} demo]
  {Synergistic Interplay between Surface Polarons and Adsorbates for Photocatalytic Nitrogen Reduction on TiO$_2$(110)}

\begin{document}


\begin{abstract}
Photocatalytic nitrogen reduction under ambient conditions represents a promising pathway toward sustainable ammonia production. However, the fundamental mechanisms, particularly the role of photogenerated charge carriers and their interactions with surface defects and adsorbates, remain elusive. Here, we employ density functional theory with Hubbard U corrections and hybrid functionals to demonstrate that the synergistic interactions between photogenerated electron polarons and point defects are essential for enabling nitrogen reduction on TiO$_2$(110). We reveal that water adsorption promotes polaron migration from subsurface to surface sites, while subsequent water dissociation stabilizes polarons near oxygen vacancies through proton coupled electron polaron transfer (PCEpT). This surface localization of polarons is critical for effective N$_2$ adsorption and activation. Our findings are consistent with previous experimental reports utilizing EPR that confirm the presence of reduced Ti species and STM, which shows the presence of water dimers on the surface. Moreover, the simultaneous interaction between polarons and reaction intermediates facilitates polaron transfer, thereby driving the completion of the nitrogen reduction reaction. Our findings elucidate the pivotal role of surface polarons in photocatalytic nitrogen fixation and provide mechanistic insights applicable to a broad range of oxide surfaces and interfaces capable of hosting small polarons, offering new design principles for efficient photocatalysts operating under ambient conditions.
\end{abstract}

\section{Introduction}
Ammonia is an essential chemical feedstock with widespread applications in the production of fertilizers, dyes, and pharmaceuticals~\cite{Smil1999,Erisman2008}. In addition to its industrial significance, ammonia has attracted increasing attention in future hydrogen economies due to its high energy density and large hydrogen content~\cite{Klerke2008}. For nearly a century, the Haber--Bosch (HB) process has been the primary industrial method for meeting global ammonia demand~\cite{Smith2020}. Although ammonia synthesis is an exothermic reaction ($\Delta H = -92$ kJ mol$^{-1}$), the HB process operates at elevated temperatures (700--850 K) and high pressures (50--200 atm) to overcome the substantial kinetic barrier associated with N$\equiv$N bond activation. These energy-intensive conditions rely heavily on nonrenewable fossil resources and contribute significantly to the global carbon footprint~\cite{Shi2019,Qian2018,Chen2018}. Consequently, the development of sustainable ammonia synthesis technologies that operate under ambient conditions has emerged as a critical long-term objective.

In this context, harnessing abundant solar energy to drive chemical reactions under ambient conditions has emerged as a promising alternative to both the Haber--Bosch process and its electrochemical counterparts~\cite{Huang2022}. Photocatalytic nitrogen reduction using transition metal oxide semiconductors, including titanium dioxide, zinc oxide, and copper oxide, has therefore been extensively investigated in experimental studies over the past decade~\cite{Medford2017}. To achieve overall sustainability, water is commonly employed as the reducing agent; however, many photocatalysts preferentially catalyze water oxidation, resulting in extremely low nitrogen reduction activity~\cite{Choe2021,Chen2020srrl}. The feasibility of ammonia synthesis from N$_2$ and water under light irradiation was first demonstrated by Schrauzer and Guth in 1977~\cite{Schrauzer1977}. Since then, numerous experimental studies have reported ammonia formation in TiO$_2$-based systems under various conditions, albeit with generally poor conversion efficiencies. Notably, Hirakawa \textit{et al.}~\cite{Hirakawa2017} achieved a maximum solar-to-chemical energy conversion efficiency of 0.02\%, which remains among the highest values reported to date. They attributed the observed activity to reduced Ti species (Ti$^{3+}$), which were proposed to serve as active sites for nitrogen fixation and to promote N$\equiv$N bond breaking as the rate-determining step~\cite{Hirakawa2017}. In contrast, Comer and Medford~\cite{Comer2018} challenged this interpretation based on \textit{ab initio} thermodynamics and the computational hydrogen electrode (CHE) approach within density functional theory, concluding that the TiO$_2$(110) surface is unlikely to be intrinsically active for nitrogen reduction due to the low stability of adsorbed N$_2$H$_x$ and NH$_x$ intermediates. Subsequently, Thiel and co-workers~\cite{Xie2019} provided further mechanistic insights into N$_2$ reduction to NH$_3$ driven by H$_2$O photolysis on TiO$_2$(110) using density functional theory calculations. Nevertheless, the detailed role of polarons and their synergistic interaction with adsorbates in driving photocatalytic nitrogen reduction has not been fully resolved.

More recent studies have established a direct link between reduced Ti species and the formation of small electron polarons in TiO$_2$~\cite{Franchini2021}. These small electron polarons, commonly described as Ti$^{3+}$(d$^1$) centers, arise from lattice polarization and local structural distortion around Ti$^{4+}$(d$^0$) sites, phenomena that are expected under typical photocatalytic conditions on the TiO$_2$(110) surface~\cite{Franchini2021}. Such polaronic states can act as catalytically relevant sites by strongly interacting with adsorbates and reaction intermediates, thereby modulating surface reactivity~\cite{Pastor2022,Reticcioli2019}. For example, Sarker \textit{et al.}\cite{PadaSarker2024} demonstrated that hole polarons located at equatorial and bridge oxygen sites can substantially lower the oxygen evolution reaction (OER) overpotential via a peroxo-type oxygen pathway, while also stabilizing key *OH, *O, and *OOH intermediates compared to polaron-free surfaces. These findings underscore the critical role of polarons in governing photocatalytic reaction energetics. Nevertheless, despite growing evidence of polaron--adsorbate coupling effects, the precise role of small polarons in water-driven photocatalytic nitrogen reduction to ammonia remains poorly understood.

In this study, we employ DFT in conjunction with the Hubbard U approach and hybrid density functionals to elucidate the formation of small electron polarons induced by surface hydroxylation and their coupled role in nitrogen reduction on the rutile TiO$_2$(110) surface. By systematically examining water adsorption and dissociation, we show that H$_2$O promotes the migration of polarons from the subsurface to the surface and stabilizes them in the vicinity of oxygen vacancy sites. The fixation of surface polarons is identified as a key factor enabling effective N$_2$ adsorption. Furthermore, polaron transfer is facilitated through their simultaneous interaction with reaction intermediates, which ultimately drives the completion of the nitrogen reduction pathway. The polaron-mediated mechanism proposed here provides detailed mechanistic insights into nitrogen reduction on TiO$_2$(110) and offers a conceptual framework for improving the sustainable photocatalytic synthesis of ammonia.

\begin{figure*}[h]
    \centering    
    \includegraphics[width=2.0\columnwidth]{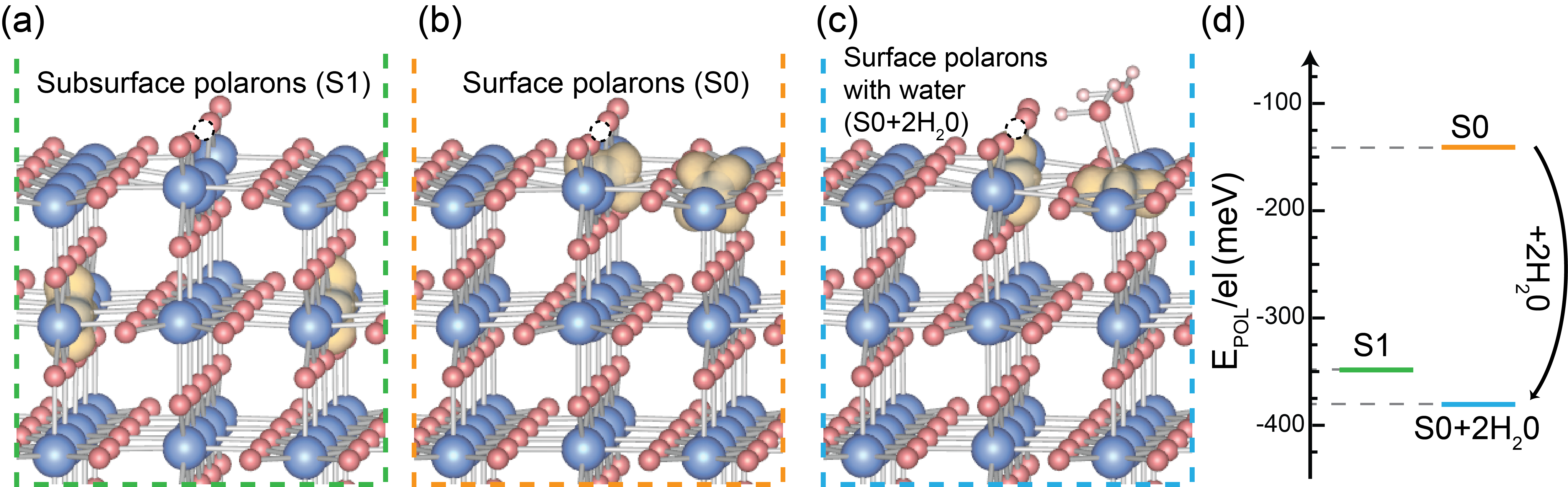}
    \caption{\textbf{Polaron formation near oxygen vacancy.} Polarons are localized at (a) subsurface (S1), (b) surface (S0), and surface with two adsorbed water (S0+2H$_2$O). (d) Calculated polaron formation energies per electron (E$_\text{POL}$/el) for all three configurations. The isosurface plot of localized electron polaron density is shown in yellow and has Ti-\textit{d} orbital nature. Isosurface value is set to 10\% of maximum. In the ball and stick model, Ti and O atoms of TiO$_2$(110) surfaces are shown in blue and red, respectively. Oxygen vacancy is shown in dashed hollow circle. Subsurface polarons are more favorable than surface polarons, whereas interplay with adsorbed water promotes the polarons to the surface and stabilizes them.} 
    \label{fig:pol}
\end{figure*}

\section{Methodology}
All calculations were performed within the framework of density functional theory (DFT) using the projector-augmented wave (PAW) pseudopotential~\cite{Blchl1994}, as implemented in the Vienna \textit{ab initio} simulation package (\texttt{VASP})~\cite{Kresse1996,Kresse1999}. Structural optimizations were carried out using the Perdew--Burke--Ernzerhof (PBE) exchange--correlation functional within the generalized gradient approximation (GGA)~\cite{Perdew1996}. The localized nature of Ti-$3d$ electrons in rutile TiO$_2$(110) was treated using the Dudarev approach~\cite{Dudarev1998}, in which a Hubbard-like on-site Coulomb correction of $U = 3.9$~eV was applied to the Ti-$d$ orbitals. This value of $U$ has been previously determined from constrained random-phase approximation calculations and has been shown to reliably describe polaron localization and defect energetics in TiO$_2$~\cite{Setvin2014}. Long-range dispersion interactions, which are important for describing adsorbate--surface interactions, were included using the Grimme DFT-D3 method~\cite{Grimme2010,Grimme2011}. Spin polarization was included in all calculations to accurately capture the formation and localization of small polaron states. The rutile TiO$_2$(110) surface was modeled using a $2 \times 3$ supercell slab comprising four stoichiometric TiO$_2$ layers (64 TiO$_2$ units, corresponding to 192 atoms in total). A vacuum spacing of 20~\AA{} was introduced along the surface normal ($c$ direction) to prevent spurious interactions between periodically repeated slabs. Brillouin zone integrations were performed using a $2 \times 2 \times 1$ Monkhorst--Pack $k$-point mesh~\cite{Monkhorst1976}. A plane-wave kinetic energy cutoff of 500~eV was employed consistently throughout all calculations. Geometry optimizations were carried out using a conjugate-gradient algorithm until the total energy change between successive ionic steps was less than $10^{-5}$~eV and the Hellmann--Feynman forces on each atom were reduced below 0.01~eV~\AA$^{-1}$. To obtain an accurate description of the electronic structure, particularly the position of defect states and band edges, electronic density of states calculations were performed using the hybrid Heyd--Scuseria--Ernzerhof (HSE06) functional~\cite{Heyd2003} with a denser $4 \times 4 \times 1$ Monkhorst--Pack $k$-point grid.

To achieve selective control over charge localization at specific Ti sites, we employed the occupation matrix control technique~\cite{Allen2014}. This approach enables the stabilization of localized electronic states by performing an initial constrained calculation in which the input occupation matrix is held fixed, thereby enforcing electron localization on a chosen Ti site. The system is then fully relaxed in a subsequent unconstrained calculation, allowing the electronic structure to evolve naturally from the imposed initial condition. This procedure ensures a reliable description of small electron polaron formation and minimizes convergence to metastable delocalized solutions.

\begin{figure*}[t]
    \centering    
    \includegraphics[width=2.0\columnwidth]{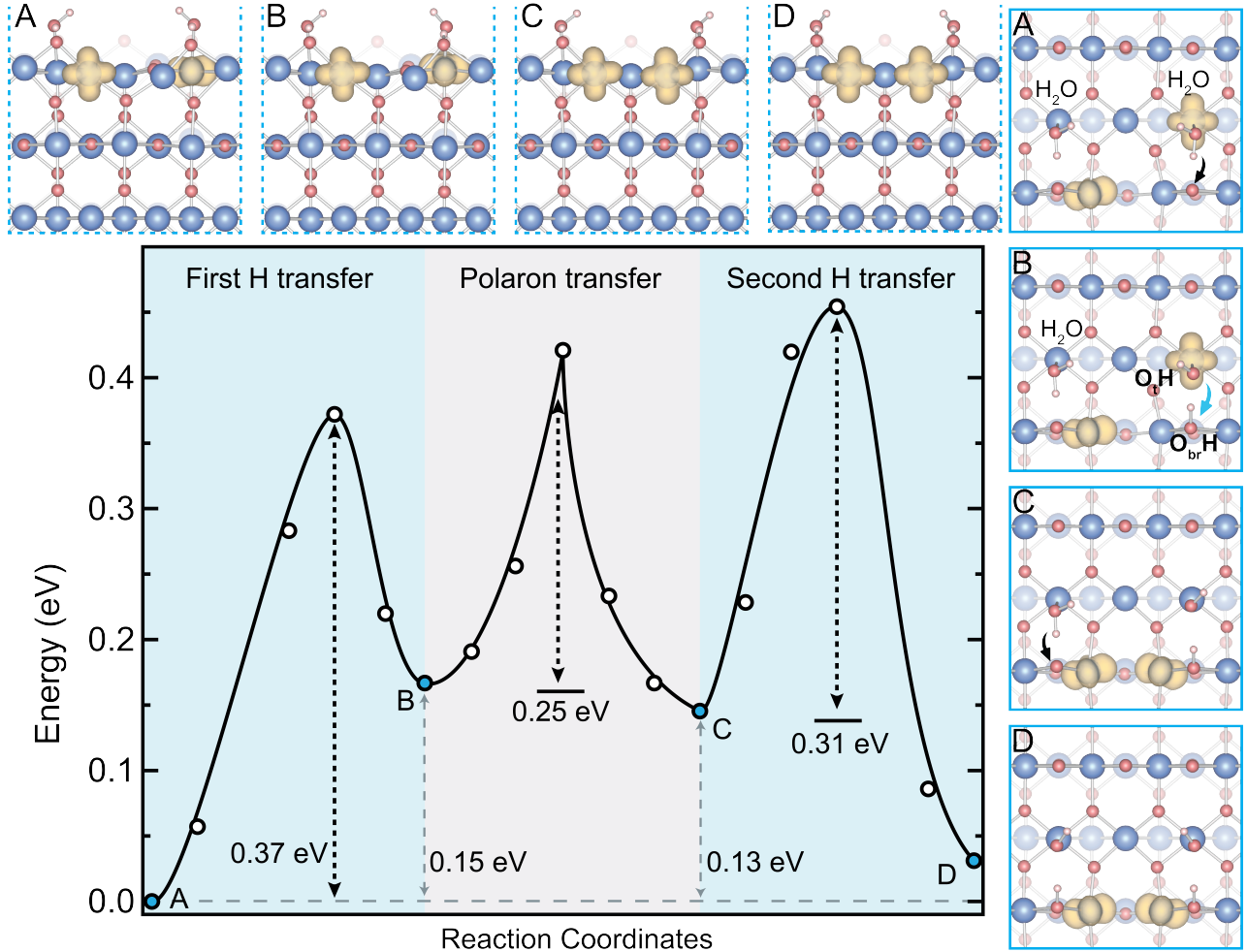}
    \caption{\textbf{Migration barriers of different steps involved in proton-coupled polaron transfer assisted by H$_2$O dissociation.} Ball-and-stick models of intermediates A, B, C, and D in the side and top view are also shown. The first H transfer happens from A to B, followed by an electron polaron transfer from B to C. The migration barrier of the second hydrogen transfer is 0.31 eV. The isosurface plot of localized electron polaron is shown in yellow. The isosurface value is set to 10\% of the maximum.} 
    \label{fig:neb}
\end{figure*}

The formation energy of a small electron polaron, defined as the energy gain associated with electron localization, was evaluated as
\begin{equation}
\begin{split}
E_{\mathrm{POL}} = E(\text{localized polaron}) \\
- E(\text{delocalized electron})
\label{polaron}
\end{split}
\end{equation}
where $E$(localized polaron) and $E$(delocalized electron) denote the total energies of the supercell containing a single excess electron in a localized polaronic state and in a delocalized state, respectively. A negative value of $E_{\mathrm{POL}}$ indicates that polaron formation is energetically favorable.

The adsorption energy of molecular nitrogen on the TiO$_2$(110) surface was calculated as
\begin{equation} 
\Delta E_{N_2} = E_{[TiO_{2}(110) + N_2]} - E_{TiO_{2}(110)} - E_{N_2} \label{ads1} 
\end{equation}
where $E_{\mathrm{TiO_2(110)} + N_2}$ is the total energy of the TiO$_2$(110) slab containing an oxygen vacancy (V$_{O}$) with an adsorbed N$_2$ molecule. $E_{\mathrm{TiO_2(110)}}$ denotes the total energy of the polaron-hosting TiO$_2$(110) slab with one V$_{O}$ in the absence of adsorbates, and $E_{N_2}$ corresponds to the total energy of an isolated N$_2$ molecule in the gas phase.

Similarly, the adsorption energies of hydrogenated nitrogen intermediates (N$_x$H$_x$) were evaluated according to
\begin{equation} 
\begin{split}
    \Delta E_{N_xH_x} = E_{[TiO_{2}(110) + N_xH_x]} - E_{TiO_{2}(110)} \\ - \frac{x}{2} E_{N_2} - \frac{x}{2} E_{H_2} 
\end{split}
\label{ads2} 
\end{equation}

where $E_{\mathrm{TiO_2(110)} + N_xH_x}$ is the total energy of the TiO$_2$(110) slab with an adsorbed N$_x$H$_x$ intermediate, $E_{\mathrm{TiO_2(110)}}$ is the total energy of the corresponding polaron-hosting slab with one oxygen vacancy (V$_{O}$), and $E_{H_2}$ represents the total energy of an isolated H$_2$ molecule in the gas phase. For the adsorption energy of *NH$_3$, the energy of a gas-phase NH$_3$ molecule was taken as the reference state.

The computational hydrogen electrode (CHE) model was employed to evaluate the Gibbs free energy of adsorption ($\Delta G$), which is defined as
\begin{equation}
\Delta G = \Delta E_{\mathrm{ad}} + \Delta \mathrm{ZPE} - T \Delta S - eU,
\label{free}
\end{equation}
where $\Delta E_{\mathrm{ad}}$ is the adsorption energy calculated using Eqs.~\ref{ads1} and \ref{ads2}, $\Delta \mathrm{ZPE}$ is the zero-point energy correction, $\Delta S$ is the entropy change, $T$ is the temperature, $e$ is the elementary charge, and $U$ denotes the applied electrode potential. Zero-point energy corrections were obtained by calculating the vibrational frequencies of adsorbed intermediates using the finite-difference method~\cite{Wu2005}. In these calculations, both the adsorbates and the coordinating surface atoms were allowed to displace along each Cartesian direction with a step size of 0.015 \AA{} in order to construct the Hessian matrix. Entropic contributions were evaluated by considering only the vibrational entropy of the adsorbed species. To elucidate the coupled mechanisms of electron polaron migration and proton transfer, activation barriers were computed using the nudged elastic band (NEB) method~\cite{Henkelman2000} as implemented in \texttt{VASP}, allowing the identification of minimum-energy pathways between initial and final states.

\section{Results and discussion}
\subsection{Formation of polarons from the surface oxygen vacancy}
The (110) termination of rutile TiO$_2$ constitutes rows of bridging oxygen (O$_\text{br}$) atoms along the [001] direction, with three-coordinated oxygen (O$_\text{3c}$) atoms located in the lower plane as shown in Figure S1 in the Supporting Information (SI). The Ti atoms situated below the row of bridging O atoms exhibit sixfold octahedral coordination (Ti$_\text{6c}$), whereas other in-plane Ti atoms have five-fold coordination (Ti$_\text{5c}$). Removal of one bridging oxygen atom leads to the creation of two five-fold coordinated Ti sites (NN-Ti$_\text{5c}$) in the nearest neighbor of the oxygen vacancy (V$_\text{O}$).

Metal oxides commonly host oxygen vacancies, as in TiO$_2$, where the removal of an oxygen atom donates two excess electrons to the system. These electrons undergo self-trapping and stabilize as small electron polarons due to the polarization potential arising from local lattice distortions~\cite{Franchini2021,Dey2021,Dey2023}. Using the occupation matrix control technique, we selectively controlled polaron localization in both the subsurface (S1) and surface (S0) regions of the TiO$_2$(110) slab. The localized polaron configurations in the S1 and S0 regions are shown in Figure~\ref{fig:pol}(a) and (b), respectively, while the associated Ti--O bond-length distortions ($D$) around each polaronic site are presented in Figure~S2 of the SI.

To quantify the energetics of polaron formation on the TiO$_2$(110) surface, we calculated the polaron formation energy per excess electron by evaluating the energy difference between localized and delocalized electronic states in the presence of an oxygen vacancy, as defined in Eq.~\ref{polaron}. The computed $E_{POL}$/el is more negative for subsurface polarons (S1: $-347$~meV) than for surface polarons (S0: $-141$~meV), indicating a thermodynamic preference for subsurface localization. Although the positively charged oxygen vacancy center (V$_\mathrm{O}^{2+}$) electrostatically attracts excess electrons, Coulombic repulsion between the polarons favors their spatial separation. Furthermore, polaron formation in the subsurface region is energetically favored due to the higher lattice distortion energy associated with surface Ti sites. These results are consistent with previous theoretical and experimental studies reporting the preferential stabilization of electron polarons in subsurface layers of TiO$_2$(110)~\cite{Setvin2014,Moses2016,Reticcioli2018}.
\begin{figure}[h!]
    \centering    \includegraphics[width=1.0\columnwidth]{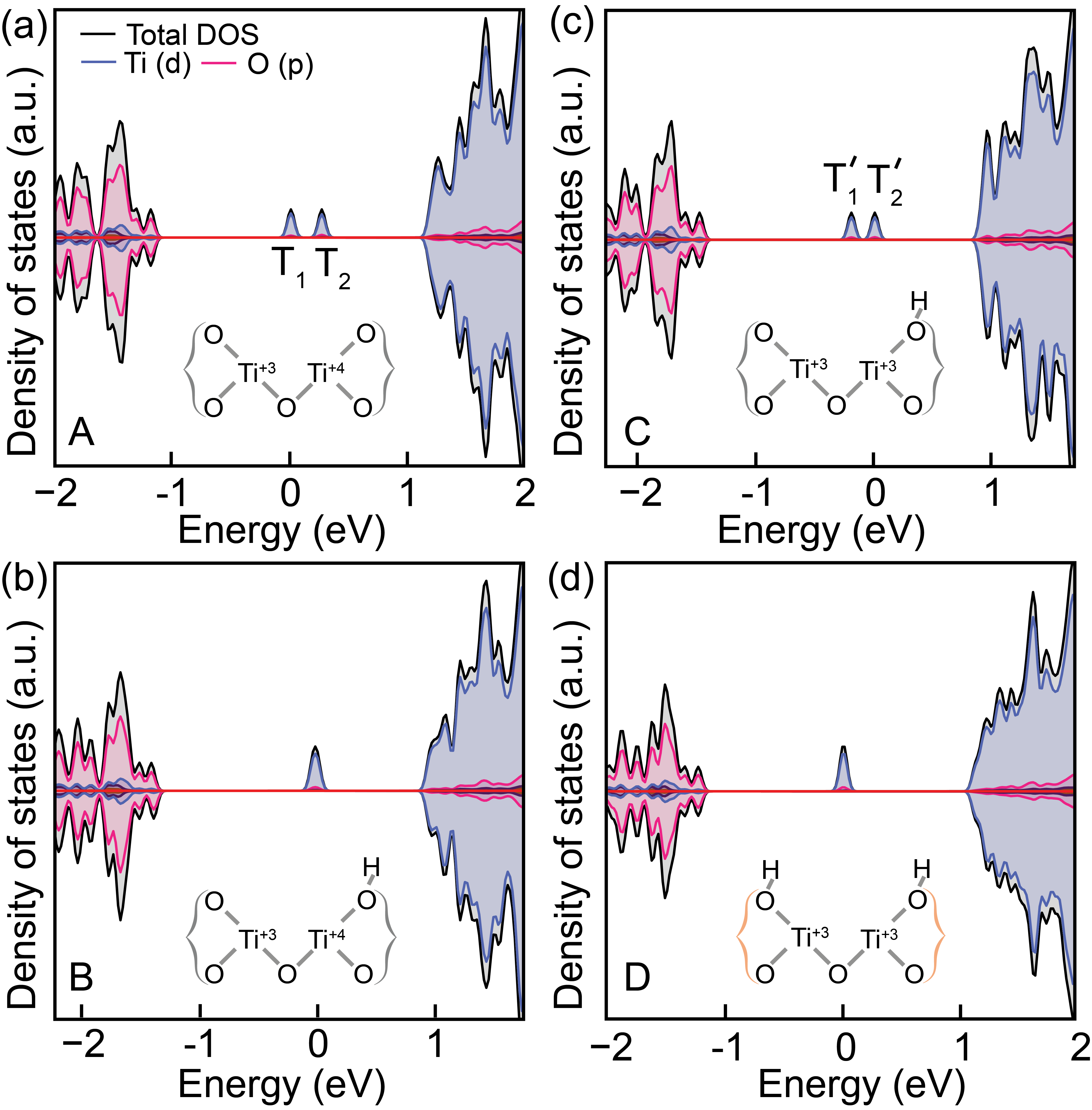}
    \caption{\textbf{Orbital-projected density of states analysis for polaron assisted H$_2$O dissociation.} (a) Upon adsorption of two water molecules (A), polarons are formed, resulting in the emergence of two quasi-degenerate in-gap states, T$_1$ and T$_2$, (b) following the transfer of the first proton, a degenerate state is formed (B), (c) the in-gap state is again split into two quasi-degenerate states (T$_\text{1}^{'}$ and T$_\text{2}^{'}$) after the polaron transfer (C), (d) the degenerate in-gap states emerges due the fixation of polarons near the oxygen vacancy (D).}
    \label{fig:wa-disso}
\end{figure}

Efficient photocatalytic nitrogen reduction requires the relocation of electron polarons from the subsurface to the surface layer of TiO$_2$(110), where they can directly participate in surface reactions. Previous studies have shown that this redistribution can be facilitated by water adsorption~\cite{Reticcioli2019,Zhu2020,Xu2022}. To explicitly investigate the effect of adsorbed water on polaron localization, we introduced two H$_2$O molecules at the nearest-neighbor Ti$_{5c}$ sites (NNN--Ti$_{5c}$) on the TiO$_2$(110) surface. The optimized structure containing surface-localized polarons in the presence of adsorbed water (denoted as S0+2H$_2$O) is shown in Figure~\ref{fig:pol}(c). In the presence of water, the formation energy of surface polarons (S0+2H$_2$O) is lowered by 238~meV relative to that of surface polarons on the clean surface (S0). Furthermore, the stability of these surface polarons is slightly enhanced by 32 meV compared to subsurface polarons (S1), as summarized in Figure~\ref{fig:pol}(d). These results demonstrate a strong coupling between adsorbed water molecules and electron polarons, which not only promotes polaron migration from the subsurface to the surface but also stabilizes surface-localized polarons, thereby creating active sites for subsequent nitrogen reduction.

\subsection{H$_2$O dissociation and transfer of polarons near the vacancy site}
Adsorbed water molecules on TiO$_2$ interact strongly with electron polarons, influencing both their stabilization and the adsorption and dissociation behavior of H$_2$O on the surface~\cite{Chen2020}. Scanning tunneling microscopy and photoemission studies show that water adsorption on rutile TiO$_2$(110) couples directly to excess electrons, drawing polarons toward the surface and stabilizing them through an attractive adsorbate--polaron interaction~\cite{Yim2018}. This coupling stabilizes polarons at the TiO$_2$(110) surface, while Coulombic repulsion still drives the two excess electrons to localize at distinct, spatially separated lattice sites. Such polaronic defects can act as active centers by regulating interfacial charge transfer and reaction energetics~\cite{Pastor2022}. Indeed, polarons on TiO$_2$(110), particularly hole polarons, have been shown to play a direct catalytic role in reactions such as CO photooxidation~\cite{Zhao2022} and the oxygen evolution reaction~\cite{PadaSarker2024}. Motivated by these findings, we propose that polarons stabilized near oxygen vacancy (V$_{O}$) sites can serve as active centers for nitrogen adsorption and activation. To elucidate this mechanism, we employed NEB calculations to resolve the stepwise migration of electron polarons in conjunction with proton transfer toward V$_{O}$ sites. We termed it as proton coupled electron polaron transfer (PCEpT) pathway and associated energy barriers are illustrated in Figure~\ref{fig:neb}.
\begin{figure*}[h!]
    \centering    \includegraphics[width=2.0\columnwidth]{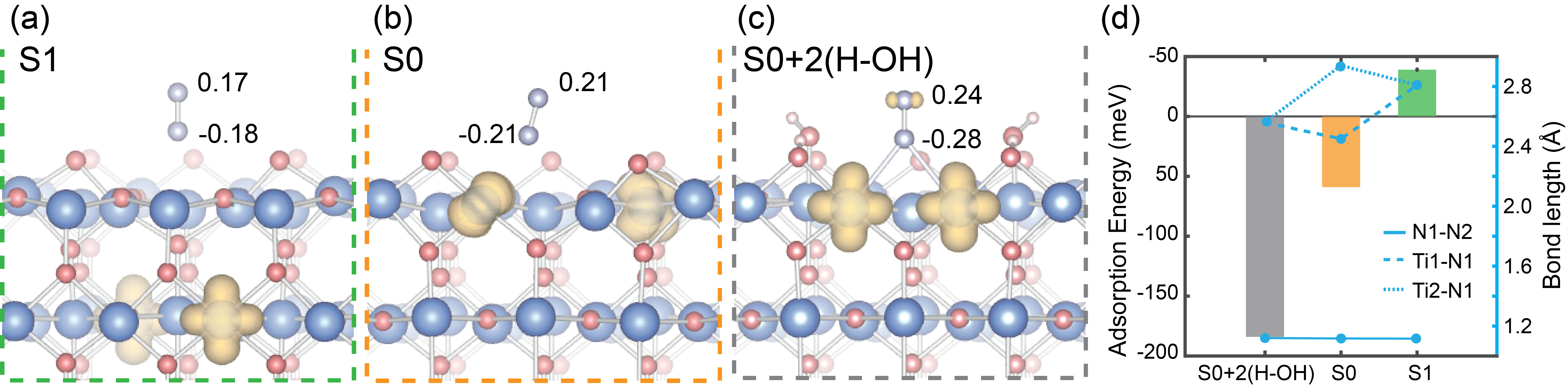}
    \caption{\textbf{Different adsorption configurations of N$_2$ on TiO$_2$(110).} Adsorbed nitrogen when two polarons (a) in the subsurface (S1), (b) in surface (S0), and attached to oxygen vacancy (S0+2(H-OH)). (d) The calculated adsorption energies along with the bond lengths. The isosurface value is set to 10\% of the maximum.}
    \label{fig:n2-ads}
\end{figure*}

In the first step, one adsorbed water molecule dissociates into --H and  $^*$OH, accompanied by proton transfer to a nearest-neighbor Ti$_{5c}$ site adjacent to the oxygen vacancy (V$_{O}$). This process leads to adsorption configuration B (Figure~\ref{fig:neb}), characterized by the formation of a bridging hydroxyl group (O$_{br}$H) at a next-nearest-neighbor Ti$_{5c}$ (NNN--Ti$_{5c}$) site and a terminal hydroxyl group (O$_{t}$H) at a nearest-neighbor Ti$_{5c}$ (NN--Ti$_{5c}$) site. The calculated activation barrier for this initial water dissociation step via proton transfer is 0.37 eV (Figure~\ref{fig:neb}). Although kinetically accessible, this step slightly destabilizes the system, with configuration B being 0.15 eV higher in energy than the initial configuration A. In the subsequent step, PCEpT occurs from the NNN--Ti$_{5c}$ site to the NN--Ti$_{5c}$ site, further reorganizing the charge distribution in the vicinity of the oxygen vacancy.

To model electron polaron transfer, we employed the Marcus/Emin--Holstein--Austin--Mott (EHAM) theory~\cite{Holstein2000,Marcus1985,Marcus1993,Austin1969}. Within this framework, the initial and final electronic states are represented by parabolic potential energy surfaces, and the transition state corresponds to their intersection. A linear interpolation scheme was used to construct the one-dimensional potential energy surfaces and to estimate the activation barrier for polaron hopping between adjacent Ti sites. The calculated migration barrier for electron polaron transfer is 0.25 eV (Figure~\ref{fig:neb}), which is slightly lower compared to the previously reported value of $\sim$0.3 eV for bulk TiO$_2$~\cite{Deskins2007}. The shape of the resulting potential energy surface indicates that electron-polaron transfer proceeds via a nonadiabatic mechanism, consistent with observations in other photocatalytic materials~\cite{Dey2021,Plata2013}. Following electron-polaron migration, the second proton transfer associated with H$_2$O dissociation occurs toward a NN--Ti$_{5c}$ site with an activation barrier of 0.31 eV (Figure~\ref{fig:neb}). This second dissociation step stabilizes the system, leading to configuration D, which features a [Ti$^{3+}$--O--Ti$^{3+}$] moiety. Such a reduced Ti pair constitutes an active site capable of strongly binding and activating N$_2$. Notably, all three activation barriers involved in the process, namely, the first proton transfer (0.37 eV), electron-polaron transfer (0.25 eV), and the second proton transfer (0.31 eV) are comparatively low. It indicates that the coupled water dissociation and polaron migration process can occur readily under ambient conditions.

To gain deeper insight into water dissociation followed by PCEpT, we analyzed the atom-projected density of states (PDOS), as shown in Figure~\ref{fig:wa-disso}. In the initial configuration containing two adsorbed water molecules and an oxygen vacancy (V$_{O}$+2H$_2$O), two quasi-degenerate polaronic trap states appear within the band gap (configuration A). These states correspond to a triplet configuration formed by two electron polarons in the majority spin channel, consistent with previous reports~\cite{Reticcioli2022,Moses2016,Cheng2021,Shibuya2012}. The lowest-energy trap state (T$_1$) lies 1.15~eV below the conduction band minimum (CBM), while the energy separation between T$_1$ and T$_2$ is 0.29~eV [Figure~\ref{fig:wa-disso}(a)], in good agreement with experimentally and theoretically reported defect levels near 1.0~eV below the CBM~\cite{Kowalski2010,Cheng2021,Moses2016}. Upon the first proton transfer to a bridging oxygen site, the two trap states collapse into a single degenerate defect level located approximately 0.90~eV below the CBM [Figure~\ref{fig:wa-disso}(b)], indicating a reorganization of the localized electronic structure. Following the subsequent electron-polaron transfer to a Ti$_{5c}$ site, two quasi-degenerate trap states reappear, with the lowest-energy state (T$_1^{\prime}$) positioned 1.02~eV below the CBM and separated from the higher-energy state by 0.19~eV [Figure~\ref{fig:wa-disso}(c)]. After the second proton transfer associated with water dissociation, these trap states again merge into a single degenerate defect level located 1.07~eV below the CBM [Figure~\ref{fig:wa-disso}(d)]. The progressive fixation of electron polarons in the vicinity of the oxygen vacancy is energetically favorable and catalytically relevant. Localized electron polarons adjacent to oxygen vacancy can serve as active sites, promoting the reaction by stabilizing charge separation and mediating the rate-limiting catalytic steps.

\subsection{N$_2$ adsorption and activation}
\begin{figure*}[h]
    \centering    \includegraphics[width=2.0\columnwidth]{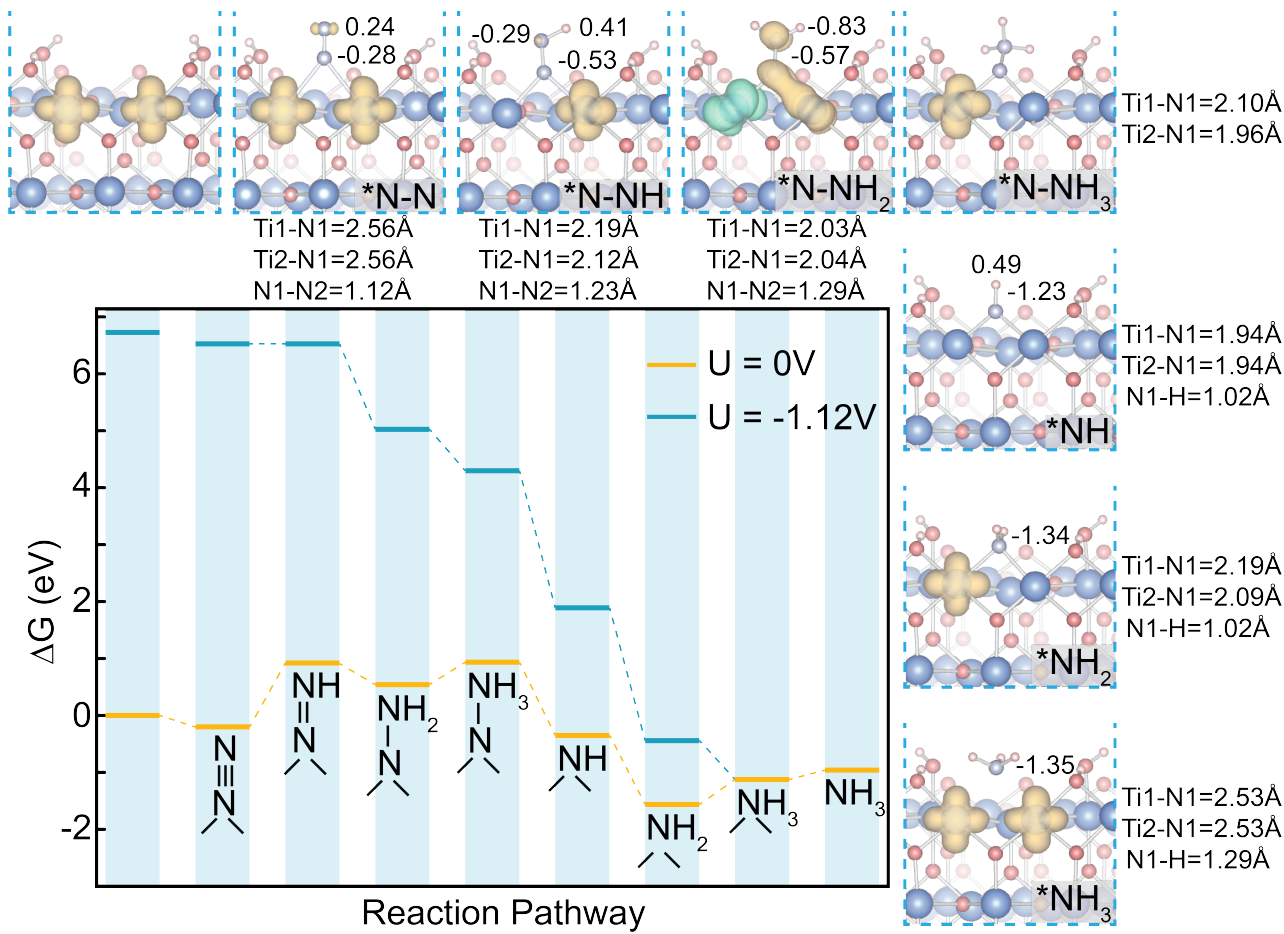}
    \caption{\textbf{Gibb's free energies of the distal mechanism of nitrogen reduction.} The simultaneous interplay of electron polarons and absorbates is shown for all intermediates in the subpanels. The calculated Bader charge transfer in units of e, is shown beside the atoms. The bond length between the left Ti site containing polaron (Ti1), and the Ti site on the right (Ti2) with lower nitrogen (N1), is illustrated, along with other crucial bond lengths of adsorbates.}
    \label{fig:free-struc}
\end{figure*}

The initial and rate-limiting step in photocatalytic nitrogen reduction toward ammonia formation is the activation of molecular dinitrogen. To examine this process, we first performed geometry optimizations of N$_2$ adsorbed on TiO$_2$(110) with electron polarons localized in the subsurface (S1), on the surface (S0), and near the surface oxygen vacancy following H$_2$O dissociation. The calculated adsorption energies obtained using Eq.~\ref{ads1} are $+39$ meV and $-59$ meV for the S1 and S0 configurations, respectively [Figure~\ref{fig:n2-ads}(d)]. The corresponding Bader charge transfer to N$_2$ is limited to 0.18$e$ (S1) and 0.21$e$ (S0), indicating weak physisorption and minimal N$_2$ activation in both cases. In contrast, when polarons are stabilized near the surface oxygen vacancy, N$_2$ binds strongly to the surface, exhibiting a significantly enhanced adsorption energy of $-184$~meV. This chemisorption is accompanied by elongation of the N--N bond to 1.12~\AA{} and a substantial charge transfer of 0.28$e$ to the adsorbed N$_2$ molecule [see Figure \ref{fig:free-struc}]. In addition, the short Ti--N bond lengths of 2.56~\AA{} in the S0+2(H--OH) configuration further indicate effective activation of the N$\equiv$N bond. Compared with previous studies on nitrogen reduction over TiO$_2$~\cite{Wu2019}, the larger charge transfer observed here underscores the importance of an explicit and accurate treatment of photogenerated charges. Following the fixation of electron polarons near V$_{O}$ through PCEpT, two localized polarons form a degenerate defect state in the middle of the band gap [Figure S3 (a) in SI]. Upon N$_2$ adsorption, this degenerate state splits into two quasi-degenerate levels separated by 0.12~eV [Figure  S3 (b) in SI], reflecting strong electronic coupling between the adsorbate and the substrate. These localized polarons, commonly referred to as bi-Ti$^{3+}$ species [or Ti$^{3+}$--O--Ti$^{3+}$ moiety], therefore act as highly active sites for N$_2$ adsorption and activation by enabling substantial charge transfer, thereby initiating the nitrogen reduction reaction.

\subsection{Reduction of adsorbed N$_2$}
The first hydrogenation step (*N$_2$ + H$^{+}$ + e$^{-}$ $\rightarrow$ *N$_2$H) is generally the most energy-demanding step in nitrogen reduction, as it requires partial weakening of the strong N$\equiv$N triple bond. Here, the calculated Gibbs free energy change for this rate-limiting step, from *N$_2$ to *N$_2$H, is 1.12~eV. During this initial protonation, one electron polaron is completely transferred from a nearest-neighbor Ti$_{5c}$ site to the adsorbed N$_2$ molecule. As a result, the N--N bond length elongates from 1.12 to 1.23~\AA{}, while the Ti--N bond distances contract to 2.12 and 2.19~\AA{}, indicating strengthened metal--adsorbate interactions. The complete transfer of an electron polaron to N$_2$ is further confirmed by the electronic density of states analysis: the low-energy polaronic defect state vanishes upon *N$_2$H formation, and two new in-gap states with opposite spin character emerge, predominantly derived from N-\textit{p} orbitals [Figure  S3 (c) in SI]. These observations demonstrate a strong synergistic interaction between the adsorbed nitrogen species and photogenerated electron polarons, which facilitates charge transfer and enables the initial activation step of the nitrogen reduction reaction.

\subsection{First ammonia release from *N-NH$_2$}
In the subsequent hydrogenation step, the second (H$^{+}$ + e$^{-}$) pair generated from water dissociation reacts with *N--NH, yielding either *NH--NH or *N--NH$_2$. Among these two pathways, the formation of *N--NH$_2$ is energetically more favorable, with a free energy change of 0.54~eV, compared to 0.72~eV for *NH--NH (Figure~S4 in SI). This energetic preference indicates that the distal mechanism is more favorable than the alternating mechanism, as it proceeds along a more downhill free-energy profile. During this step, pronounced charge redistribution occurs between the adsorbate and the TiO$_2$ surface, accompanied by electron back-donation. As a consequence, a new electron polaron with opposite spin character is formed, while another polaronic charge becomes partially delocalized between a Ti$_{5c}$ site and the nitrogen atom. This behavior is reflected in the electronic density of states of the *N--NH$_2$ intermediate, which exhibits distinct in-gap states [Figure S3 (d) in SI]. The up-spin in-gap state originates primarily from hybridized N-\textit{p} and Ti-\textit{3d} orbitals, whereas the down-spin in-gap state is dominated by Ti-\textit{3d} character. The redistribution of photogenerated charges further weakens the N--N bond, leading to an elongation from 1.23 to 1.29~\AA{} as shown in Figure \ref{fig:free-struc}. Concurrently, stronger metal--adsorbate interactions are established, as evidenced by the shortening of the Ti--N bond lengths to 2.03 and 2.04~\AA{} for Ti1 and Ti2, respectively.

Subsequently, the interaction of the third (H$^{+}$ + e$^{-}$) pair with the *N--NH$_2$ intermediate leads to the formation and release of the first ammonia molecule, leaving an adsorbed *N species on the surface (*N--NH$_2$ + H$^{+}$ + e$^{-}$ $\rightarrow$ *NH$_3$ + *N). This transformation is accompanied by the localization of an electron polaron at a Ti$_{5c}$ site. The associated polaronic trap state is dominated by Ti $3d$ orbital character and is located 0.94~eV below the CBM, as shown in Figure S3 (e) in SI.

\subsection{Second ammonia release via successive protonation}
Subsequently, the electron polaron is fully transferred to the adsorbed *N species, accompanied by proton transfer from photogenerated H$^{+}$ originating from the dissociation of H$_2$O molecules coordinated to nitrogen, resulting in the formation of *N--H. During the following hydrogenation step, the transfer of an additional H$^{+}$ proton to *N--H induces the localization of another electron polaron at a Ti$_{5c}$ site, giving rise to a new polaronic trap state located 0.97~eV below the CBM. Upon completion of the subsequent hydrogenation and release of the second NH$_3$ molecule, two excess photogenerated electrons localize in the vicinity of the oxygen vacancy (V$_{O}$). These localized polarons form two quasi-degenerate defect states within the band gap, separated by 0.37~eV. Notably, the Gibbs free energy associated with NH$_3$ desorption ($\Delta G{_{des}}$) is only 0.16~eV, indicating facile ammonia release, as this barrier is lower than the thermal energy available under ambient conditions. Overall, the synergistic interplay between photogenerated polarons, adsorbed intermediates, and oxygen vacancies enables efficient charge donation and back-donation processes, thereby promoting the nitrogen reduction reaction through successive polaron-mediated electron transfer steps.
\begin{figure}[h]
    \centering   \includegraphics[width=1.0\columnwidth]{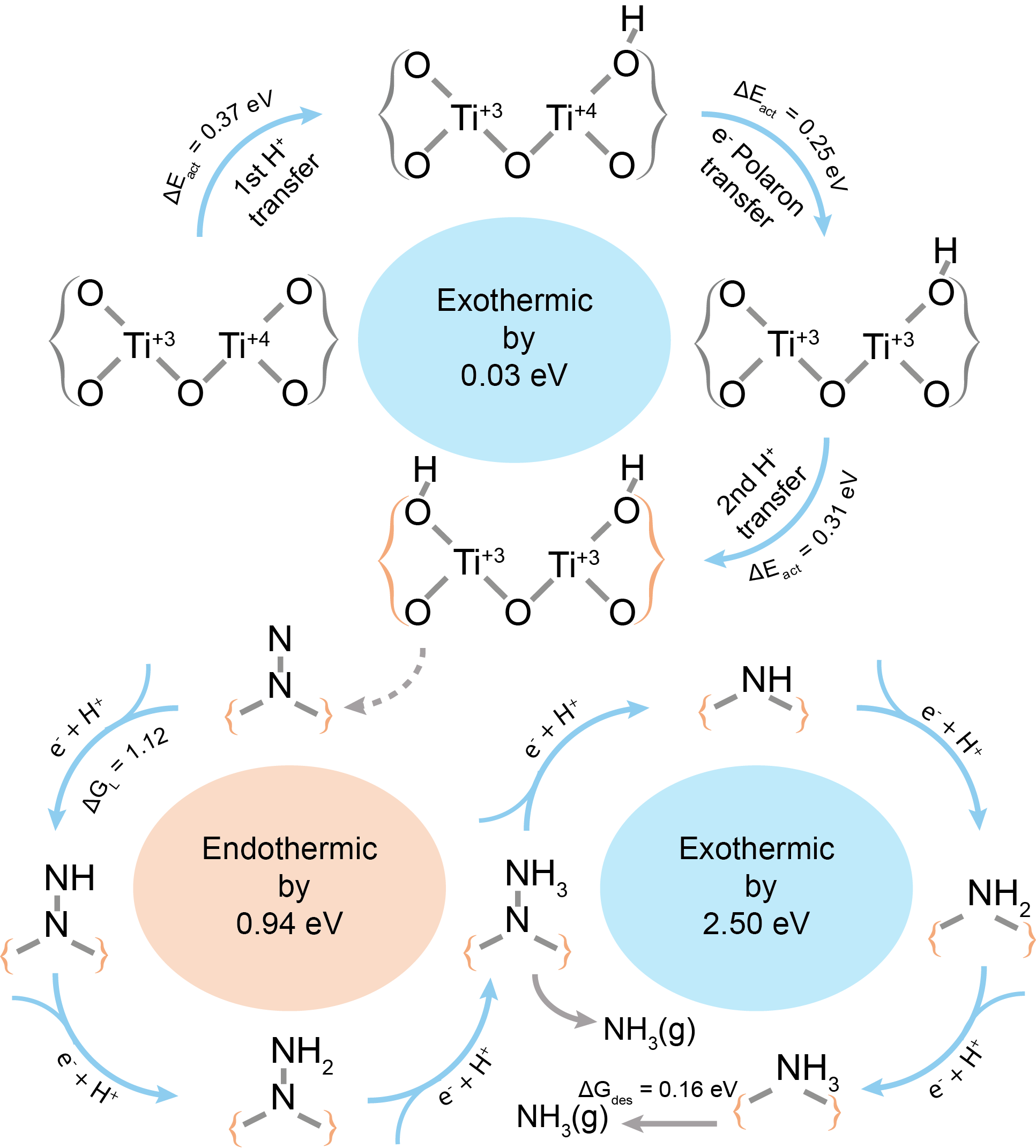}
    \caption{\textbf{Complete proposed mechanism for photocalytic nitrogen reduction on TiO$_2$(110).} A schematic illustrating mechanism for hydroxylation on the TiO$_2$(110) surface and photocatalytic nitrogen reduction.}
    \label{fig:mechanism}
\end{figure}

\subsection{Discussion}
The complete photocatalytic nitrogen reduction mechanism on the TiO$_2$(110) surface is summarized schematically in Figure~\ref{fig:mechanism}. Surface hydroxylation induced by water dissociation, together with the fixation of electron polarons near the V$_{O}$ site, is thermodynamically favorable, resulting in an overall exothermicity of 0.03~eV. This process involves three charge-transfer events that require activation by photoexcitation. Specifically, electron-polaron migration to a neighboring Ti$^{4+}$ site proceeds via a nonadiabatic barrier of 0.25~eV, while the two associated proton-transfer steps exhibit activation energies of 0.37 and 0.31~eV, respectively. Following nitrogen activation through the distal pathway, the release of the first NH$_3$ molecule is overall endothermic by 0.94~eV. Notably, this energy requirement is lower than the limiting free energy for the initial protonation of N$_2$ (1.12~eV), highlighting the beneficial role of polaron-mediated charge transfer. The subsequent formation and desorption of the second NH$_3$ molecule is strongly exothermic, with a net energy release of 2.50~eV, while the NH$_3$ desorption free energy is only 0.16~eV, indicating facile product release under ambient conditions. The band edge alignement to the vacuum level are carried out using electrostatic potential alignment from slab calculations (details in the Section S7 in SI). Owing to the presence of an oxygen vacancy, two localized polarons, and dissociated water species, a strong surface dipole is formed, which substantially lowers the electron affinity and decreases the work function. Similar decrease in work function upon removal of surface oxygen has also been observed experimentally in UPS measurements of rutile TiO$_2$~\cite{Borodin2011}. Referencing the conduction-band minimum to the vacuum level yields $-0.86$~eV, corresponding to a reducing power of 3.63~eV relative to the absolute nitrogen reduction potential ($-4.49$~eV~\cite{Comer2018}). Consequently, the resulting photocatalytic overpotential becomes strongly negative ($-2.51$~eV), indicating that nitrogen reduction at such defect-engineered TiO$_2$ surfaces is thermodynamically downhill. Together, these results elucidate the energetic landscape governing surface hydroxylation, polaron dynamics, and nitrogen reduction on TiO$_2$(110), and provide a comprehensive mechanistic framework for understanding and optimizing photocatalytic ammonia synthesis.

Recent XPS measurements show that defects in rutile TiO$_2$ act as efficient charge--trapping centers that nearly double photocatalytic activity by stabilizing photogenerated electrons rather than promoting recombination~\cite{Wagstaffe2020}. Our simulations corroborate this behavior, demonstrating that the trapped charges can be efficiently transferred to reactant species. The presence of water dimers on rutile TiO$_2$(110) has also been directly observed using STM~\cite{Yim2018}. Furthermore, Hirakawa \textit{et al.}~\cite{Hirakawa2017} confirmed N$_2$ adsorption on Ti$^{3+}$ species using electron spin resonance (ESR) spectroscopy. Measurements performed at 77~K under vacuum revealed a characteristic signal at $g = 2.004$, attributed to bridging oxygen vacancies. Similar strong EPR signals at $g = 1.95$ have also been reported and assigned to reduced Ti sites adjacent to V$_\mathrm{O}$~\cite{Zhang2020}. Our study resolves that this triplet state originates from a bipolaron bound to a V$_\mathrm{O}$, often described as a bi-Ti$^{3+}$ species or Ti$^{3+}$--O--Ti$^{3+}$ moiety, which serves as the active site. We further elucidate the detailed mechanism by which surface polarons interact with adsorbates and mediate catalytic reactions. Furthermore, metal loading with Fe, Ni, and Co has also been shown to promote the formation of oxygen vacancies and Ti$^{3+}$ species due to altered coordination environments, thereby enhancing the catalytic activity of rutile TiO$_2$~\cite{Chen2021}. Our approach, which explicitly accounts for polarons, oxygen vacancies, and their synergistic interactions with adsorbates, can be extended to understand such effects more broadly. More generally, adsorbate--polaron interactions in transition metal oxides are critical for the rational design of improved TMO photocatalysts.

\section{Conclusion}
\noindent
In summary, we have elucidated the critical role of surface-localized electron polarons and their interactions with defects and adsorbates in enabling photocatalytic nitrogen reduction on TiO$_2$(110). Water adsorption and dissociation drive polaron migration from subsurface to surface sites and stabilize them near oxygen vacancies, a process essential for initiating N$_2$ adsorption and activation. Furthermore, the simultaneous interaction between polarons and reaction intermediates governs polaron transfer and facilitates the complete reduction of N$_2$ to NH$_3$. Our results provide a detailed mechanistic picture of polaron localization, migration, and charge transfer throughout the reaction pathway. Beyond establishing fundamental insights into polaron-mediated catalysis on titania, the computational methodologies and mechanistic principles developed here are broadly applicable to other oxide surfaces and interfaces capable of hosting small polarons. These findings establish new design principles for developing efficient, sustainable, and cost-effective photocatalysts for ambient ammonia synthesis, with implications extending to other challenging reduction reactions in renewable energy conversion and green chemistry.

\begin{acknowledgement}
The authors thank the Materials Research Centre (MRC), and Supercomputer Education and Research Centre (SERC), Indian Institute of Science, Bangalore, for providing computational facilities. The authors acknowledge support from the Institute of Eminence (IoE) scheme of The Ministry of Human Resource Development, Government of India.
\end{acknowledgement}

\begin{suppinfo}
The Supporting Information is available free of charge
\begin{itemize}
  \item Atomic structures of the TiO$_2$(110) surface in side and top views, including surface, subsurface, and lower subsurface layers and oxygen vacancy sites; total energies, zero-point energies, and vibrational entropic corrections of all reference gas-phase molecules; average Ti–O bond-length distortions associated with polaron localization at different surface and subsurface sites; adsorption free energies of all nitrogen reduction reaction intermediates with thermodynamic corrections; orbital-projected density of states of adsorbed intermediates; spin-density distributions of localized polarons; and comparative free-energy, spin-density, and electronic structure analyses of distal and alternating reaction; potential alignment and reducing power   .
\end{itemize}
\end{suppinfo}

\providecommand{\latin}[1]{#1}
\makeatletter
\providecommand{\doi}
  {\begingroup\let\do\@makeother\dospecials
  \catcode`\{=1 \catcode`\}=2 \doi@aux}
\providecommand{\doi@aux}[1]{\endgroup\texttt{#1}}
\makeatother
\providecommand*\mcitethebibliography{\thebibliography}
\csname @ifundefined\endcsname{endmcitethebibliography}
  {\let\endmcitethebibliography\endthebibliography}{}

\end{document}


\title{Supporting Information for:\\ ``\textit{Synergistic Interplay between Surface Polarons and Adsorbates for Photocatalytic Nitrogen Reduction on TiO$_2$(110)}''}

\author {Manoj Dey}
\author {Ritesh Kumar}
\author {Abhishek Kumar Singh\thanks{abhishek@iisc.ac.in}}
\affiliation{Materials Research Centre, Indian Institute of Science, Bangalore-560012, India}

\date{\today}

\maketitle

\tableofcontents

\newpage
\section{Section S1: Atomic structure of TiO$_2$(110) surface}

\begin{figure}[h!]
    \centering
    \includegraphics[width=0.6\columnwidth]{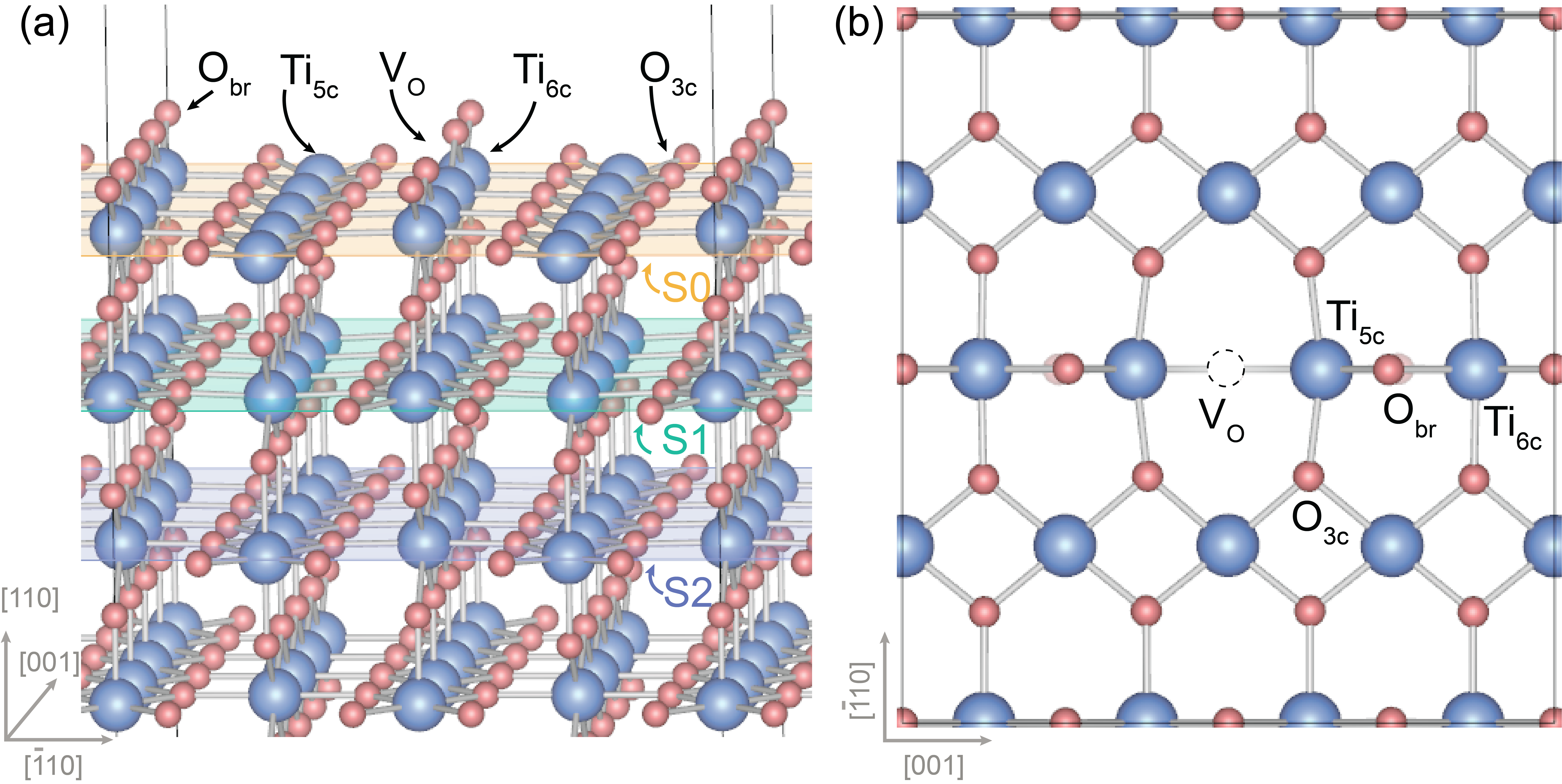}
    \caption{Atomic structure of TiO$_2$(110) surface in (a) side view, and (b) top view. Ti atoms, O atoms, and O vacancy are shown in blue balls, red balls, and hollow circles, respectively. The upper surface (S0), sub-surface (S1), and lower subsurface (S2) are shown in orange, green, and blue planes, respectively.}
    \label{fig:mechanism}
\end{figure}

\section{Section S2: Energies of the reference molecules}
Total energies (E$_\text{total}$), zero point energies (ZPE), and vibrational entropic corrections (TS) of all the reference molecules are listed below

\begin{table}[h]
\centering
\begin{tabular}{|l|c|c|c|}
\hline
Reference Molecule & E\textsubscript{total} (eV) & ZPE (eV) & TS (eV) \\
\hline
H\textsubscript{2} & -6.76 & 0.27 & 0.41 \\
N\textsubscript{2} & -16.62 & 0.15 & 0.59 \\
NH\textsubscript{3} & -19.73 & 0.91 & 0.60 \\
\hline
\end{tabular}
\end{table}

\clearpage
\section{Section S3: Average bond-length distortion}
In order to quantify the lattice distortion around polaron localization, we have calculated the average bond length distortion (D) as,
\begin{equation}
D = 1/n_O \sum_{i = 1,n_O} |\Delta B_i|
\end{equation}
where, $n_O$ is the number of O atoms, and {$\Delta$ B$_\text{i}$} is the Ti-O bond length change around the polaron configuration in comparison to the pristine. The calculated average bond-length distortion (D) for both polarons, when formed in the subsurface (S1) is 0.074 \AA. D is small (0.051 \AA) when polaron is the nearest neighbor (NN-Ti$_\text{5c}$) to oxygen vacancy, while the average bond length distortion is large (0.109 \AA) in surface Ti$_\text{5c}$ site.
\begin{figure}[h!]
    \centering
    \includegraphics[width=0.6\columnwidth]{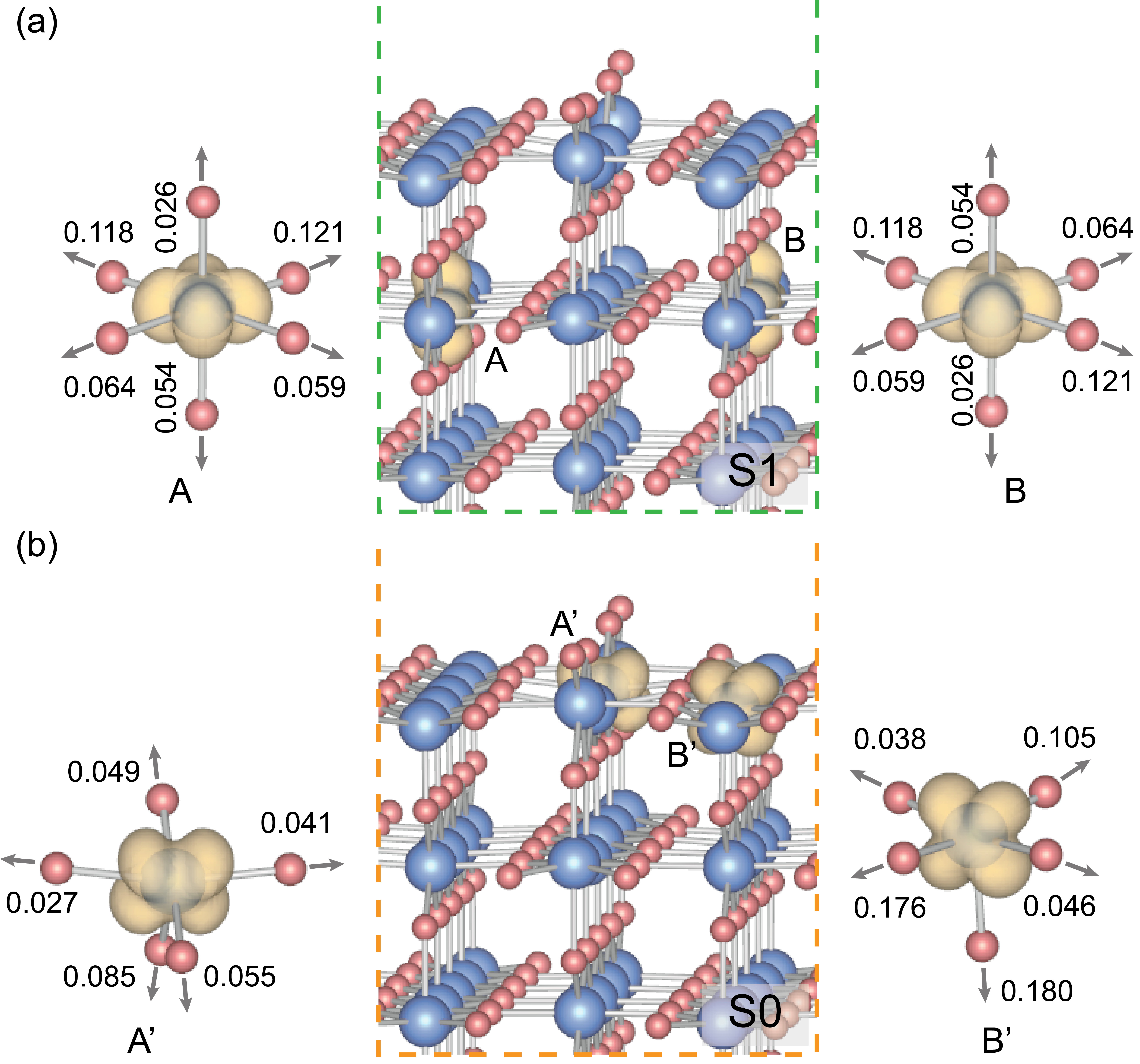}
    \caption{The polarons localization and corresponding bond length distortions when polarons form in (a) subsurface (S1), (b) surface (S0).}
    \label{fig:mechanism}
\end{figure}

\clearpage
\section{Section S4: Adsorption free energies of all intermediates}
\begin{table}[h]
\centering
\begin{tabular}{|l|c|c|c|c|}
\hline
Species & E$_{ad}$ (eV) & ZPE-TS (eV)& G\textsubscript{0} (eV)& G\textsubscript{U} (eV)\\
\hline
\hline
N$_{2}$ (g) & 0.000 & 0.000 & 0.000 & 6.7268 \\
*N$_{2}$ & -0.1838 & -0.01639 & -0.2002 & 6.5266 \\
*N-NH & 0.5677 & 0.3532 & 0.9209 & 6.5266 \\
*N-NH\textsubscript{2} & -0.1294 & 0.6700 & 0.5406 & 5.0251 \\
*N-NH\textsubscript{3} & -0.1359 & 1.0708 & 0.9349 & 4.2983 \\
*NH & -0.6369 & 0.2854 & -0.3515 & 1.8908 \\
*NH\textsubscript{2} & -1.9721 & 0.4084 & -1.5637 & -0.4426 \\
*NH\textsubscript{3} & -1.9103 & 0.7855 & -1.1249 & -1.1249 \\
\hline
\end{tabular}
\end{table}

\section{Section S5: Density of states of intermediates}

\begin{figure*}[h]
    \centering    \includegraphics[width=1.0\columnwidth]{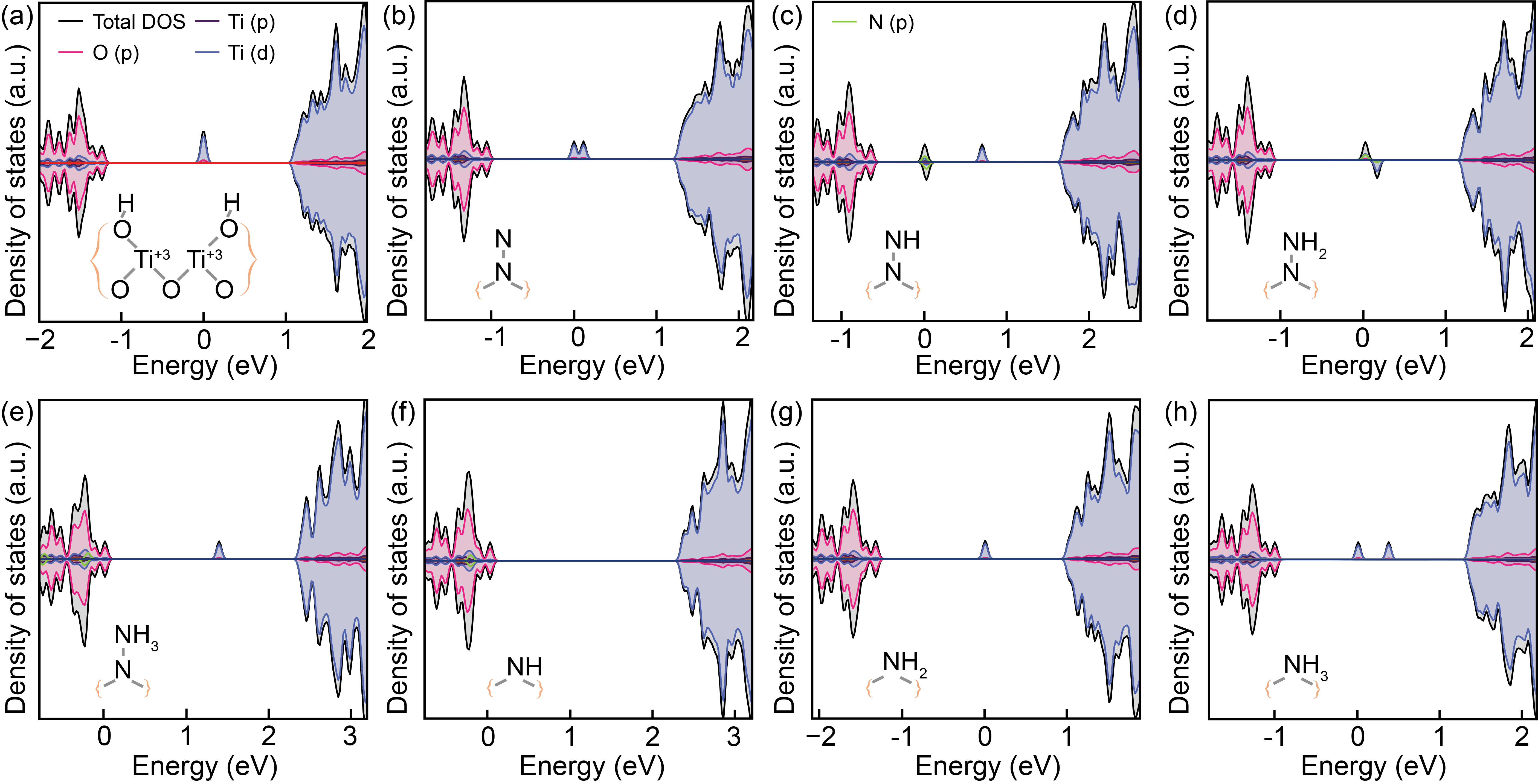}
    \caption{Orbital projected density of states analysis of intermediates. Orbital-project density of states of all the intermediates of the nitrogen reduction reaction. The contribution of Ti-\textit{d}, O-\textit{p}, and N-\textit{p} orbitals are shown in blue, red, and green colors, respectively.}
    \label{fig:int-dos}
\end{figure*}

\clearpage

\section{Section S6: Distal vs Alternating}
\begin{figure}[h!]
    \centering
    \includegraphics[width=0.5\columnwidth]{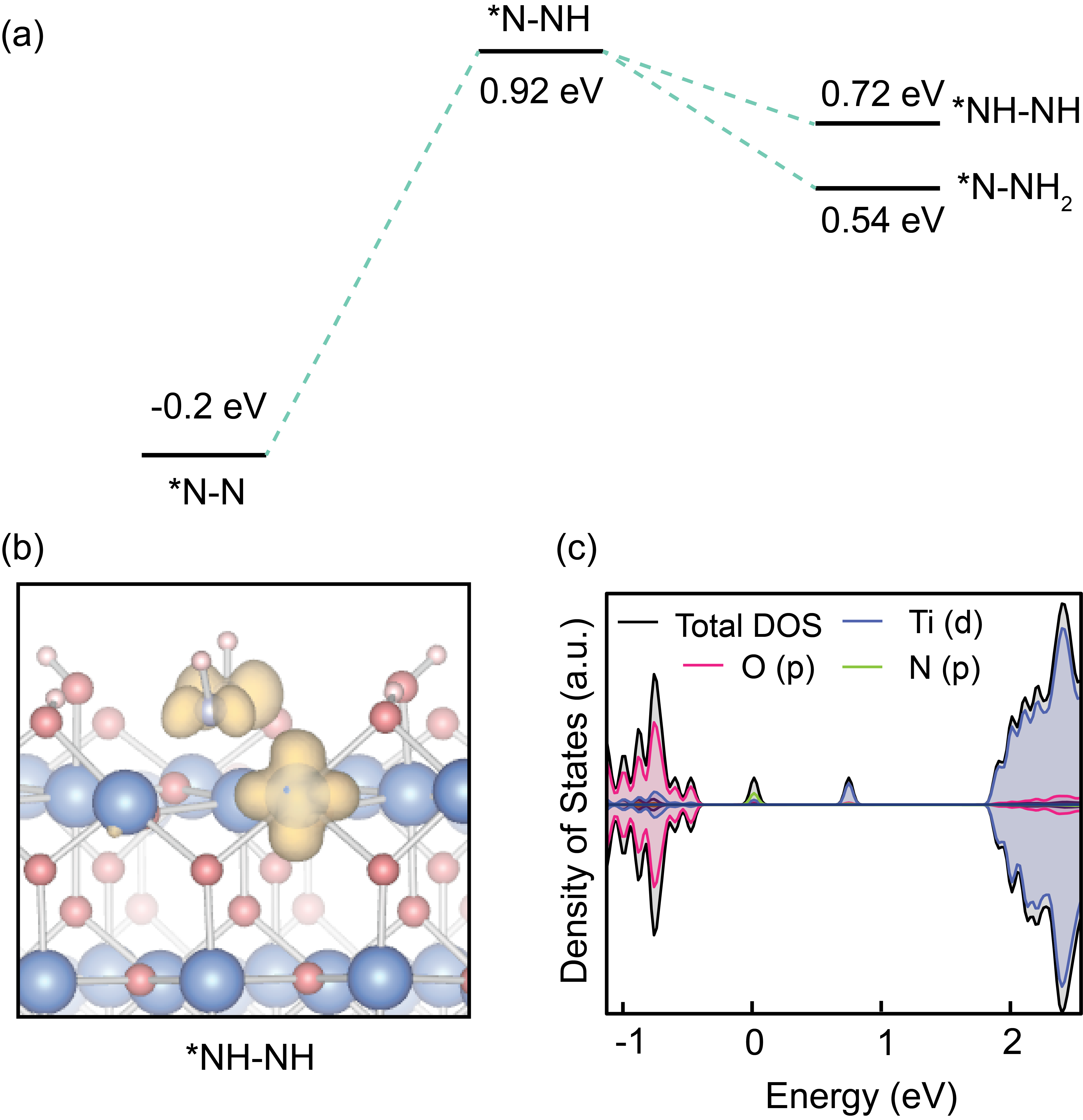}
    \caption{(a) Free energy comparison of possible distal vs alternating mechanism, (b) spin-density of polaron localization, and (c) orbital-projected density of states.}
    \label{fig:mechanism}
\end{figure}

\section{Section S7: Vacuum level alignment and reducing power}
The absolute alignment of the electronic band edges was obtained using electrostatic potential referencing to the vacuum level. Slab calculations were performed for the TiO$_2$(110) surface including oxygen vacancies, localized polarons, and dissociated water, and the planar-averaged electrostatic potential was extracted using HSE06 functional. Alignment to the vacuum level was then achieved according to
\begin{equation}
E_{VBM/CBM}^{\mathrm{vac}} = E_{VBM/CBM} - E_{\mathrm{vac}},
\end{equation}
The vacuum level $E_{\mathrm{vac}}$ is the electrostatic potential in the vacuum region.

\begin{table}[h]
\centering
\caption{Vacuum-level alignment and derived energetic quantities for defective and hydrated TiO$_2$(110). All energies are given in eV.}
\label{tab:band_alignment}
\begin{tabular}{l c}
\hline\hline
Quantity & Value (eV) \\
\hline
Vacuum level, $E_{\mathrm{vac}}$ & 4.67 \\
Valence-band maximum, $E_{\mathrm{VBM}}$ & 0.7349 \\
Conduction-band minimum, $E_{\mathrm{CBM}}$ & 3.8081 \\
Band gap, $E_g$ & 3.0732 \\
Vacuum-aligned valence-band maximum, $E_{\mathrm{VBM}}^{\mathrm{vac}}$ & $-3.94$ \\
Vacuum-aligned conduction-band minimum, $E_{\mathrm{CBM}}^{\mathrm{vac}}$ & $-0.86$ \\
Nitrogen reduction potential (vacuum scale) & $-4.49$ \\
Reducing power & 3.63 \\
Photocatalytic overpotential & $-2.51$ \\
\hline\hline
\end{tabular}
\end{table}

The reducing power was defined as the energy difference between the vacuum-aligned conduction-band minimum and the absolute nitrogen reduction potential, taken as $-4.49$~eV on the vacuum scale. The photocatalytic overpotential was subsequently evaluated as the difference between the thermodynamic reaction barrier (1.12~eV) and the calculated reducing power. All numerical values resulting from this alignment are summarized in Table~\ref{tab:band_alignment}.

Using the alignment procedure described above, the vacuum level of the defective and hydrated TiO$_2$(110) slab was determined to be 4.67~eV. The vacuum-aligned valence- and conduction-band edges were obtained as $E_{\mathrm{VBM}}^{\mathrm{vac}} = -3.94$~eV and $E_{\mathrm{CBM}}^{\mathrm{vac}} = -0.86$~eV, respectively. Alignment of the conduction-band minimum to the absolute nitrogen reduction potential ($-4.49$~eV) gives a reducing power of 3.63~eV. With a Gibbs free energy barrier of 1.12~eV for the rate-limiting step ($\Delta G_{L}$), the corresponding photocatalytic overpotential ($\Delta G_{L}$ - reducing power) is calculated to be $-2.51$~eV.
